\documentclass[aps,pra,twocolumn,superscriptaddress]{revtex4-1}

\usepackage{amsmath,amssymb,amsfonts}
\usepackage{graphicx}
\usepackage{siunitx}
\usepackage{natbib}
\usepackage{braket}

\usepackage{url}
\usepackage[breaklinks]{hyperref}
\hypersetup{
    unicode=true,
    colorlinks=true,
    linkcolor=blue,
    citecolor=blue,
    urlcolor=blue
  }
	\usepackage{breakurl}

\graphicspath{ {Figures/} }

\begin{document}

\title{
Interspecies thermalization in an ultracold mixture of Cs and Yb in an optical
trap}

\author{A. Guttridge}
\email{alexander.guttridge@durham.ac.uk}
 \affiliation{Joint Quantum Centre (JQC) Durham-Newcastle, Department of Physics, Durham University, South Road, Durham, DH1 3LE, United Kingdom.}
\author{S. A. Hopkins}
\affiliation{Joint Quantum Centre (JQC) Durham-Newcastle, Department of Physics, Durham University, South Road, Durham, DH1 3LE, United Kingdom.}
\author{S. L. Kemp}
\affiliation{Joint Quantum Centre (JQC) Durham-Newcastle, Department of Physics, Durham University, South Road, Durham, DH1 3LE, United Kingdom.}
\author{Matthew D. Frye}
\email{matthew.frye@durham.ac.uk} \affiliation{Joint Quantum Centre (JQC)
Durham-Newcastle, Department of Chemistry, Durham University, South Road,
Durham, DH1 3LE, United Kingdom.}
\author{Jeremy~M.~Hutson}
\affiliation{Joint Quantum Centre (JQC) Durham-Newcastle, Department of Chemistry, Durham University, South Road, Durham, DH1 3LE, United Kingdom.}
\author{Simon L. Cornish}
\email{s.l.cornish@durham.ac.uk}
\affiliation{Joint Quantum Centre (JQC) Durham-Newcastle, Department of Physics, Durham University, South Road, Durham, DH1 3LE, United Kingdom.}

\begin{abstract}
We present measurements of interspecies thermalization between ultracold
samples of $^{133}$Cs and either $^{174}$Yb or $^{170}$Yb. The two species are
trapped in a far-off-resonance optical dipole trap and $^{133}$Cs is
sympathetically cooled by Yb. We extract effective interspecies thermalization
cross sections by fitting the thermalization measurements to a kinetic model,
giving \mbox{$\sigma_{\mathrm{Cs^{174}Yb}} = \left(5 \pm 2\right) \times
10^{-13} \, \mathrm{cm^{2}}$} and \mbox{$\sigma_{\mathrm{Cs^{170}Yb}} =
\left(18 \pm 8\right) \times 10^{-13} \, \mathrm{cm^{2}}$}. We perform quantum
scattering calculations of the thermalization cross sections and optimize the
CsYb interaction potential to reproduce the measurements. We predict scattering
lengths for all isotopic combinations of Cs and Yb. We also demonstrate the
independent production of $^{174}$Yb and $^{133}$Cs Bose-Einstein condensates
using the same optical dipole trap, an important step towards the realization
of a quantum-degenerate mixture of the two species.
\end{abstract}

\date{\today}

\maketitle

The realization of ultracold atomic mixtures
\cite{Modugno2001,Mudrich2002,Hadzibabic2002,Taglieber2008,Spiegelhalder2009,
Taie2010,McCarron2011,Ridinger2011,Wacker2015,Vaidya2015,Grobner2016,Flores2017}
has opened up the possibility of exploring new regimes of few- and many-body
physics. Such mixtures have been used to study Efimov physics
\cite{Tung2014,Pires2014,Maier2015}, probe impurities in Bose gases
\cite{Scelle2013}, and entropically cool gases confined in an optical lattice
\cite{Catani2009}. Pairs of atoms in the mixtures can be combined using
magnetically or optically tunable Feshbach resonances to create ultracold
molecules
\cite{Koehler2006,Ni2008,Lang2008,Aikawa2009,Koeppinger2014,Molony2014,Takekoshi2014,Park2015,Guo2016}.
These ultracold molecules have a wealth of applications, such as tests of
fundamental physics \cite{Isaev2010,Flambaum2007,Hudson2011}, realization of
novel phase transitions \cite{Micheli2007,Potter2010,Goral2002}, and the study
of ultracold chemistry \cite{Ospelkaus2010,Krems2008}. In addition, the
long-range dipole-dipole interactions present between pairs of polar molecules
make them useful in the study of dipolar quantum matter
\cite{Santos2000,Lahaye2009} and ultracold molecules confined in an optical
lattice can simulate a variety of condensed-matter systems
\cite{Barnett2006,Gorshkov2011,Yan2013}.

Although the large majority of work on ultracold molecules has focused on
bi-alkali systems, there is burgeoning interest in pairing alkali-metal atoms
with divalent atoms such as Yb
\cite{Nemitz2009,Tassy2010,Ivanov2011,Hansen2011,Hara2011,Baumer2011} or Sr
\cite{Pasquiou2013}. The heteronuclear $^{2}\Sigma$ molecules formed in these
systems have both an electric and a magnetic dipole moment in the ground
electronic state. The extra magnetic degree of freedom opens up new
possibilities for simulating a range of Hamiltonians for spins interacting on a
lattice and for topologically protected quantum information processing
\cite{Micheli2006}.

One of the challenging aspects of creating molecules in these systems is that
the Feshbach resonances tend to be narrow and sparse. They are
narrow because the main coupling responsible for them is the weak distance
dependence of the alkali-metal hyperfine coupling, caused by the spin-singlet
atom at short range \cite{Zuchowski2010}. They are sparse because only
molecular states with the same value of the alkali-metal magnetic quantum
number $M_F$ as the incoming atomic channel can cause resonances. The resonance
positions are determined by the (often unknown) background scattering length
\cite{Zuchowski2010,Brue2013}, and for some systems may be at impractically
high magnetic fields. Amongst the various alkali-Yb combinations, CsYb has
been proposed as the most favorable candidate because the high mass of Cs
facilitates a higher density of bound states near threshold and its large
hyperfine coupling constant increases the resonance widths \cite{Brue2013}.
However, the short-range part of the molecular ground-state potential is not
known accurately enough to predict background scattering lengths, so
experimental characterization is essential before accurate predictions of
resonance positions and widths can be made.

Here we present simultaneous optical trapping of Cs and Yb and first
measurements of the scattering properties of $^{133}$Cs+$^{174}$Yb and
$^{133}$Cs+$^{170}$Yb. We measure interspecies thermalization in the optical
dipole trap and use a kinetic model to extract effective thermalization cross
sections. We model these cross sections using quantum scattering calculations,
taking full account of the anisotropy of differential cross sections and
thermal averaging. We obtain an optimized interaction potential and use it to
make predictions of the scattering lengths for all accessible isotopologs. For
all isotopes except $^{176}$Yb, binary quantum-degenerate mixtures of Cs and Yb
are expected to be miscible at fields around 22~G, where the Efimov minimum in
the three-body recombination rate allows efficient evaporation of Cs to quantum
degeneracy \cite{Kraemer2006}.

\begin{figure}
	\centering
		\includegraphics[width=1\linewidth]{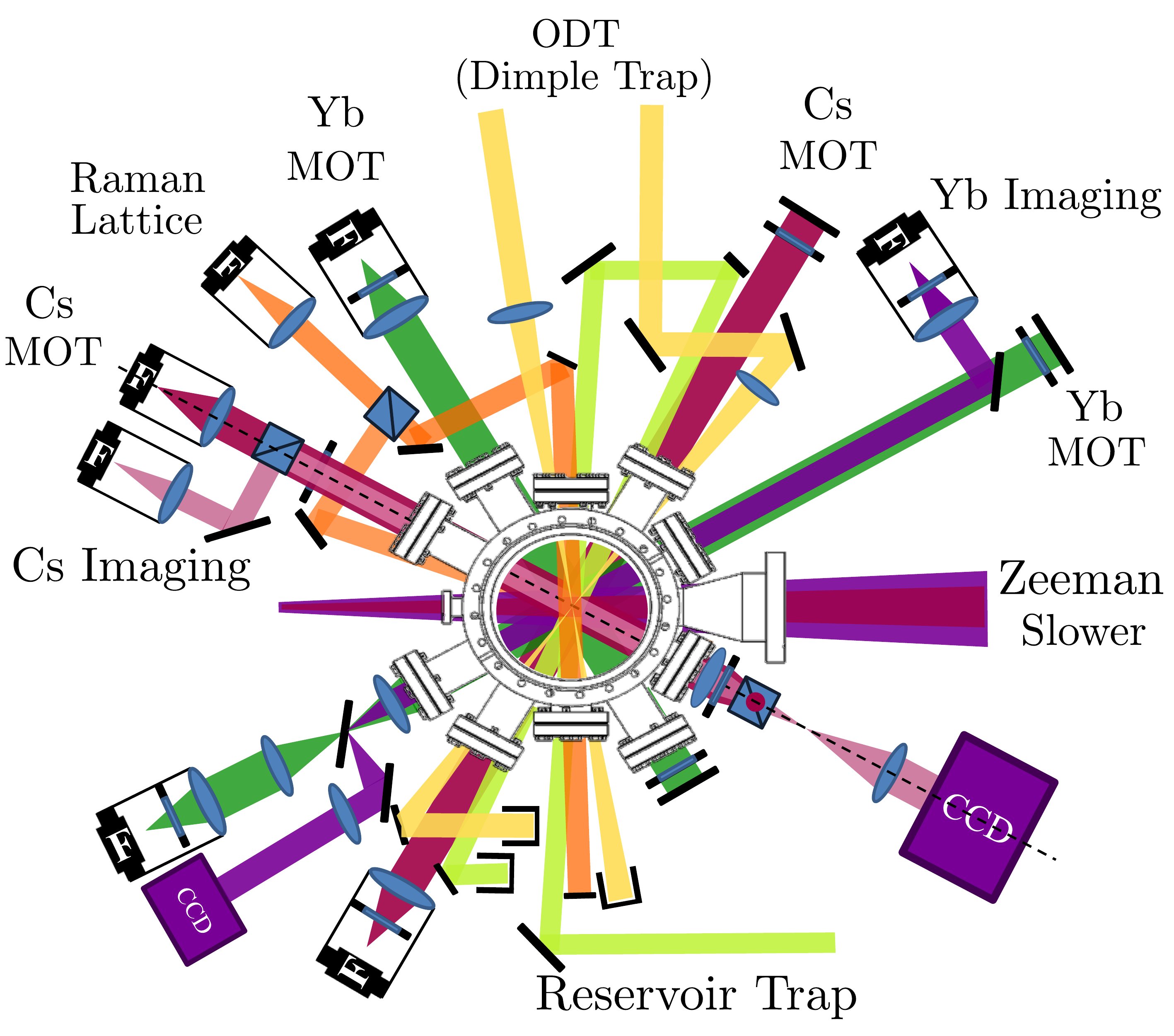}
\caption{Optical layout of the science chamber in the horizontal plane. The Cs
(Yb) imaging beam is combined with the Cs (Yb) MOT beam using a polarizing beam
splitter (dichroic mirror) and then separated after the chamber and aligned
onto a CCD camera. The Raman lattice beams used for DRSC are split using a
polarizing beam splitter, with one arm retro-reflected and the other arm dumped
after the first pass. The ODT used for thermalization measurements is referred
to as the ``dimple trap" to distinguish it from the large-volume ``reservoir
trap'' that is used for the preparation of Cs BEC.}
	\label{fig:optics}
\end{figure}

\section{Experiment}
A detailed description of our experimental apparatus can be found in
\cite{Kemp2016}, but we will summarize the main components here. Cs and Yb
magneto-optical traps (MOTs) are sequentially loaded from an atomic beam that
effuses from a dual-species oven and is slowed by a dual-species Zeeman slower
\cite{Hopkins2016}. The Cs atomic beam is slowed and trapped in the MOT using
the $^{2}S_{1/2}\rightarrow {^{2}P_{3/2}}$ transition at $\lambda = 852 \,
\mathrm{nm}$. For Yb we use the broad $^{1}S_{0}\rightarrow {^{1}P_{1}}$
transition at $\lambda = 399 \, \mathrm{nm}$ ($\Gamma / 2\pi = 29 \,
\mathrm{MHz}$) for Zeeman slowing and absorption imaging, and the narrow
$^{1}S_{0}\rightarrow {^{3}P_{1}}$ transition at $\lambda = 556  \,
\mathrm{nm}$ ($\Gamma / 2\pi = 182 \, \mathrm{kHz}$) for laser cooling in the
MOT. The optical layout of our science chamber is shown in Fig.\
\ref{fig:optics}.

The thermalization measurements presented here take place in an optical dipole
trap (ODT) formed from the output of a broadband fiber laser (IPG YLR-100-LP)
with a wavelength of \mbox{$1070 \pm 3$ nm}. The ODT consists of two beams
crossed at an angle of 40$^{\circ}$ with waists of $33 \pm 3 \,\mu$m and $72
\pm 4 \, \mu$m respectively. The intensity of each beam is independently
controlled by a water-cooled acousto-optic modulator. Yb has a moderately low
polarizability at the trapping wavelength ($\alpha_{\mathrm{Yb}}\left(1070 \,
\mathrm{nm}\right) =  150 \, a_{0}^{3}$) so that, for the powers used in the
thermalization measurements, Yb atoms are trapped only in the part of the
potential where the axial confinement is provided by the second ODT beam. Cs,
on the other hand, has a much larger polarizability at the trapping wavelength
($\alpha_{\mathrm{Cs}}\left(1070 \, \mathrm{nm}\right) = 1140 \, a_{0}^{3}$),
creating a trap deep enough that Cs atoms are confined both inside and outside
the crossed-beam region of the ODT. Some Cs atoms thus experience a trapping
potential dominated by just a single ODT beam.

\begin{figure*}
	\centering
		\includegraphics[width=0.95\linewidth]{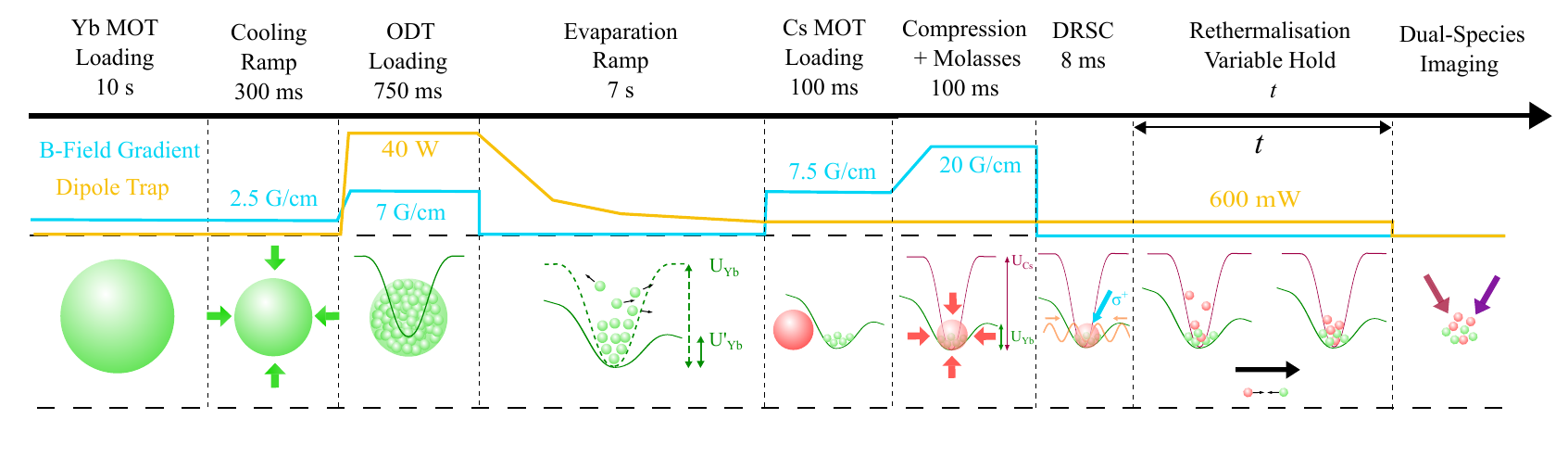}
\caption{Simplified experimental sequence. The Yb MOT is loaded, then cooled and
compressed to facilitate subsequent loading into an ODT. The Yb is then
evaporated in the ODT by ramping the trap depth until a temperature of $T = 550
\, \mathrm{n K}$ is reached. The displaced Cs MOT is loaded before it is
compressed, cooled and transferred into a near-detuned optical lattice for
DRSC. The DRSC stage loads Cs into the ODT, where it is held with Yb for a
variable time $t$ before the trap is switched off and the atoms are
destructively imaged after a variable time of flight using dual-species
absorption imaging.}
	\label{fig:sequence}
\end{figure*}

A summary of the experimental sequence used for the thermalization measurements
is shown in \mbox{Fig.\ \ref{fig:sequence}}. The two species are sequentially
loaded into the dipole trap to avoid unfavorable inelastic losses from
overlapping MOTs \footnote{We observe a significant drop in the Cs number when
the Cs MOT is operated in the presence of the \mbox{399 nm} Yb Zeeman slower
light. This drop in number is likely due to ionization of Cs atoms in the $6 \,
^{2}P_{3/2}$ state}. We choose to prepare the Yb first due to the much longer
loading time of the MOT and its insensitivity to magnetic fields. We first load
the Yb MOT for 10 s, preparing $5 \times 10^{8}$ atoms at $T= 140 \, \mathrm{\mu
K}$ \cite{Guttridge2016}, before ramping the power and detuning the MOT beams
to cool the atoms to $T= 40 \, \mathrm{\mu K}$. We load $1.8 \times 10^{7}$
atoms into the ODT with a trap depth of $U_{\mathrm{Yb}}=950 \, \mathrm{\mu
K}$. We then evaporatively cool the atoms by exponentially reducing the trap
depth to $U_{\mathrm{Yb}}=5 \, \mathrm{\mu K}$ in $7 \,$s, producing a sample
of $1 \times 10^{6}$ Yb atoms at a temperature of $T = 550 \, \mathrm{n K}$. At
this stage the Yb trap frequencies as measured by center-of-mass oscillations
are $240 \, \mathrm{Hz}$ radially and $40 \, \mathrm{Hz}$ axially.

Once the Yb is prepared in the dipole trap, the Cs MOT is loaded for \mbox{0.15
s}, at which point the MOT contains $1 \times 10^{7}$ atoms. The Cs MOT is then
compressed via ramps to the magnetic field, laser intensity and detuning before
it is overlapped with the ODT using magnetic bias coils.  The Cs atoms are then
further cooled by optical molasses before transfer into a near-detuned lattice
with $P = 100 \, \mathrm{mW}$, where the atoms are then polarized in the
$\ket{F=3,m_{F}=+3}$ state and cooled to $T = 2 \, \mathrm{\mu K}$ with 8 ms of
degenerate Raman sideband cooling (DRSC). During this stage $9 \times 10^{4}$
atoms are transferred into the ODT and the magnetic bias field is set to
\mbox{22.3 G}, corresponding to the Efimov minimum in the Cs three-body
recombination rate \cite{Kraemer2006}. During the transfer the atoms are heated
to $T = 5 \, \mathrm{\mu K}$. The heating and poor efficiency of the transfer
into the ODT are due to the poor mode matching of the DRSC-cooled cloud and the
deep ODT ($U_{\mathrm{Cs}}= 85 \, \mathrm{\mu K}$). This huge ratio of trap
depths $U_{\mathrm{Cs}}/U_{\mathrm{Yb}}= 15.5$ is greater than the (still
large) ratio of the polarizabilities
$\alpha_{\mathrm{Cs}}/\alpha_{\mathrm{Yb}}= 7.2$ due to the effect of gravity
on the weak Yb trap. The ratio of the mean trap frequencies between the two
species is $\overline{\omega}_{\mathrm{Cs}} / \overline{\omega}_{\mathrm{Yb}} =
3.1$.

The thermalization measurements thus begin with a mixture of $1 \times 10^{6}$
Yb atoms in their spin-singlet ground state $^{1}S_{0}$ and $9 \times 10^{4}$
Cs atoms in their absolute ground state $^{2}S_{1/2} \, \ket{3,+3}$. For each
experimental run the number and temperature are determined by quickly turning
off the ODT after a variable hold time and performing resonant absorption
imaging of both species after a variable time of flight.

\begin{figure*}
	\centering
		\includegraphics[width=0.95\linewidth]{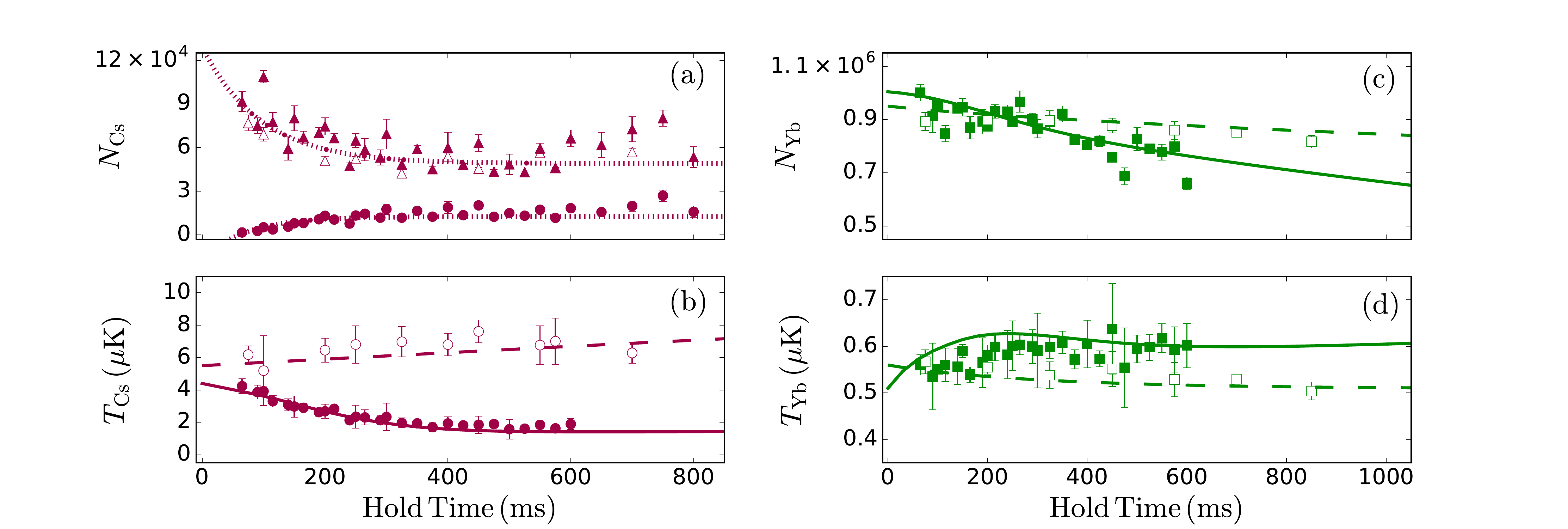} \caption{Results
of thermalization experiments. (a) and (b) show the evolution of the Cs number
and temperature as a function of hold time $t$. (c) and (d) show the $^{174}$Yb
number and temperature as a function of the same hold time. Filled symbols
indicate the presence of both Cs and $^{174}$Yb in the ODT, whereas open symbols
indicate the presence of only one species in the trap. For the Cs number,
triangles indicate the number in the single-beam region of the trap and circles
the number in the crossed-beam region, while dotted lines show the
interpolating functions used to constrain the Cs number in the model. The
dashed line shows the result of our kinetic model with only one species trapped
and the solid line shows the result for the two-component mixture. }
	\label{fig:CsYb}
\end{figure*}

\mbox{Figure \ref{fig:CsYb}} shows the number and temperature evolution of Cs
and $^{174}$Yb atoms, with and without the other species present. The smaller
initial number of Cs atoms is chosen to reduce the density of Cs such that
the effects of three-body recombination play a relatively small role in the
thermalization \footnote{We find that a larger initial density of Cs atoms
results in a larger final temperature difference between the two species and a
greater loss of Cs atoms during the thermalization}. Treatment of the number
evolution of the Cs atoms requires careful attention due to the presence of Cs
atoms both in the crossed-beam region of the trap and in the wings, where
confinement is due to only a single ODT beam. Although the Cs atoms in the
crossed- and single-beam regions are in thermal equilibrium, the atoms have
different density distributions due to the different potentials experienced.
This is an important effect to consider when calculating the spatial overlap of
the Cs and Yb atoms. We observe an increase in the number of Cs atoms trapped
in the crossed-beam region of the trap in the presence of Yb, which we
attribute to interspecies collisions aiding the loading of this region. We do
not observe any Cs atoms loaded into the crossed-beam portion of the trap in
the absence of Yb, so the number is not plotted in this case. For the Cs atoms
in the single-beam region, we estimate the axial trapping frequency to be the
same as for a single-beam trap, $5 \, \mathrm{Hz}$, and the radial frequencies
to be the same as in the crossed-beam region.

We observe a decay of the Yb number throughout the thermalization. The
timescale of this decay is much shorter than the single-species $1/e$ background
lifetime of \mbox{15 s} and we attribute the number loss to sympathetic
evaporation \cite{Mudrich2002}. The small change in the Yb temperature is
explained in part by the evaporation of hotter atoms and also by the large
number ratio $N_{\mathrm{Yb}}/N_{\mathrm{Cs}}$, which causes the final mean
temperature of the sample to be close to the initial Yb temperature. In
contrast to Yb, we observe a large change in the temperature of the Cs atoms
for short times due to elastic collisions with the Yb atoms. However, for
longer times we see the two species reach a steady state at two distinct
temperatures. The higher final temperature for Cs results from a Cs heating
rate that balances the sympathetic cooling rate.

\section{Rate equations for thermalization}

To model the thermalization results, we formulate a set of coupled equations
that describe the number and temperature kinetics. We expand upon the usual
treatment \cite{Mosk2001,Anderlini2005,Tassy2010,Ivanov2011} by including terms
for evaporation \cite{Ketterle1996} and single-species three-body recombination
\cite{Weber2003} as described in the Appendix. The coupled equations for the
number $N_{i}$ and temperature $T_{i}$ of the two species
are
\begin{widetext}
\begin{align}
\dot{N}_{i}&=
-N_{i}\gamma_{ii}\eta_i\exp(-\eta_i)
-K_{\mathrm{bg}} N_{i}
-K_{i,3}\left<n_{i}^{2}\right>_{\rm sp} N_{i},
\label{Eq:Number} \\
\dot{T}_{i}&=
\eta_i \exp(-\eta_i)\gamma_{ii}\left(1-\frac{\eta_i + \kappa_i}{3}\right)T_{i}
+K_{i,3}\left<n_{i}^{2}\right>_{\rm sp}\frac{(T_{i}+T_{i,\mathrm{H}})}{3}
\pm \frac{\xi \Gamma_{\mathrm{CsYb}} \Delta T \left(t\right)}{3N_{i}}
+ \dot{T}_{i,\mathrm{ODT}},
\label{Eq:Temp}
\end{align}
\end{widetext}
where $i=\left\{\mathrm{Yb,Cs}\right\}$, $\eta_i=U_{i}/k_{\mathrm{B}}T_{i}$,
and $\kappa_i = \left(\eta_i - 5\right)/\left(\eta_i-4\right)$
\cite{OHara2001}. $K_{\mathrm{bg}}$ is the background loss rate, $K_{i,3}$ is
the three-body loss coefficient, $n_{i}(\bf{r})$ is the density and
$\left<\ldots\right>_{\rm sp}$ represents a spatial average. $T_{i,\mathrm{H}}$
is the recombination heating term, which accounts for the increase in
temperature due to the release of the molecular binding energy during
recombination \cite{Weber2003}. We choose to neglect the three-body loss
coefficient for Yb, $K_{\mathrm{Yb},3}$, because we do not observe any evidence
of three-body loss on the experimental timescale in single-species Yb
experiments. The Cs three-body loss coefficient is measured to be
$K_{\mathrm{Cs},3}=1^{+1}_{-0.9} \times 10^{-26} \, \mathrm{cm^{6}/s}$ at the
bias field used in the measurements. In addition to the above terms,
$\dot{T}_{i,\mathrm{ODT}}$ is added as an independent heating term to account
for any heating from the trapping potential, such as off-resonant photon
scattering \footnote{Estimated to be $60 \, \mathrm{nK/s}$ for Cs using our
trap parameters} or additional heating effects due to the multi-mode nature of
the trapping laser \cite{Sofikitis2011,Menegatti2013,Lauber2011,Yamashita2017}.
The heating rate for Yb alone is found to be zero within experimental error, so
$\dot{T}_{{\rm Yb,ODT}}$ is fixed at 1 nK/s, which is the predicted heating
rate due to off-resonant photon scattering. Eq.\ \ref{Eq:Temp} uses the fact
that the average energy transferred in a hard-sphere collision is $\xi
k_{\mathrm{B}} \Delta T$, where $\xi =
4m_{\mathrm{Cs}}m_{\mathrm{Yb}}/\left(m_{\mathrm{Cs}}+m_{\mathrm{Yb}}\right)^{2}$,
$m_i$ is the mass of species $i$, and $\Delta
T=T_{\mathrm{Cs}}-T_{\mathrm{Yb}}$.

The effective intraspecies collision rate per atom for thermalization is
$\gamma_{ii}=\left<n_{i}\right>_{\rm sp}\sigma_{ii}\bar{v}_{ii}$, where
$\sigma_{ii}$ is an effective energy-independent scattering cross section. In a
hard-sphere model, the effective total interspecies collision rate is
$\Gamma_{\mathrm{CsYb}}=\bar{n}_{\mathrm{CsYb}} \sigma_{\mathrm{CsYb}}
\bar{v}_{\mathrm{CsYb}}$, where the mean thermal velocity $\bar{v}_{ij}$ is
\begin{equation}
\bar{v}_{ij}
=\sqrt{\frac{8 k_{\mathrm{B}}}{\pi}\left(\frac{T_{i}}{m_{i}}+\frac{T_{j}}{m_{j}}\right)}
\end{equation}
and the spatial overlap $\bar{n}_{\mathrm{CsYb}}$ is found by integrating the
density distributions of the two species,
\begin{widetext}
\begin{align}
\bar{n}_{\mathrm{CsYb}}&= \int \left[n_{\mathrm{Cs, single}}\left(\bf{r}\right)
+n_{\mathrm{Cs,cross}}\left(\bf{r}\right)\right]n_{\mathrm{Yb}}\left(\bf{r}\right) d^{3}r \nonumber \\
&=N_{\mathrm{Yb}}\frac{m_{\mathrm{Yb}}^{3/2} \overline{\omega}_{\mathrm{Yb}}^{3}}
{2 \pi k_{\mathrm{B}}}\left[\frac{N_{\mathrm{Cs,single}}}
{\left(T_{\mathrm{Yb}}+\beta_{\mathrm{single}}^{-2} T_{\mathrm{Cs}}\right)^{3/2}}
+\frac{N_{\mathrm{Cs,cross}}}{\left(T_{\mathrm{Yb}}
+\beta_{\mathrm{cross}}^{-2} T_{\mathrm{Cs}}\right)^{3/2}}\right]. \label{eq:interspecies density}
\end{align}
\end{widetext}
Here $\overline{\omega}_{\mathrm{Yb}}
=\sqrt[3]{\omega_{x}\omega_{y}\omega_{z}}$ is the mean Yb trap frequency and
$\beta_{j}^{2}$ is defined by $\beta_{j}^2
m_{\mathrm{Yb}}\overline{\omega}_{\mathrm{Yb}}^{2}
=m_{\mathrm{Cs}}\overline{\omega}_{\mathrm{Cs},j}^{2}$. Here
$j=\left\{\mathrm{single,cross}\right\}$ denotes the different cases for Cs
atoms trapped in the crossed- and single-beam regions.

Due to the large difference in trapping potentials between the two species, Yb
experiences a greater gravitational sag than the tightly trapped Cs. For the
case of two clouds spatially separated by $\Delta z$, the spatial overlap must
be reduced by a factor $F_{z}\left(\Delta z \right)$, where
\begin{equation}
F_{z} \left(\Delta z\right) = \exp\left(-\frac{m_{\mathrm{Yb}}\omega_{\mathrm{Yb},z}^{2}\Delta z^{2}}{2 k_{\mathrm{B}} \left(T_{\mathrm{Yb}}+\beta^{-2}_{\rm cross} T_{\mathrm{Cs}}\right)}\right) . \label{eq:correction}
\end{equation}

\section{Analysis of Results}

The coupled equations (\ref{Eq:Number}) and (\ref{Eq:Temp}) are solved
numerically. We perform least-squares fits to the experimental results to
obtain optimal values of the parameters $\sigma_{\mathrm{CsYb}}$,
$T_{\mathrm{Cs,H}}$ and $\dot{T}_{\mathrm{Cs, ODT}}$. The solid lines in
\mbox{Fig.\ \ref{fig:CsYb}} show the results of the fitted model, while the
dashed lines show the results in the absence of interspecies collisions. Fig.\
\ref{fig:CsYb}(a) does not include model results, because our analysis does not
include the kinetics of Cs atoms entering and leaving the crossed-beam region.
We instead constrain the number of Cs atoms inside and outside this region
using interpolating functions (dotted lines in figure) matched to the
experimentally measured values.

Since the origin of the heating present on long timescales is unknown, we
initially fitted both $T_{\mathrm{Cs,H}}$ and $\dot{T}_{\mathrm{Cs, ODT}}$. We
found that these two parameters are strongly correlated, with a correlation
coefficient of 0.99 \cite{Hughes2010}. We therefore choose to extract the
parameter $\dot{T}_{\mathrm{Cs, Heat}}$ corresponding to the total heating rate
from both recombination heating and heating due to the ODT. As shown in
\mbox{Fig.\ \ref{fig:CsYb}}, the best fit, corresponding to
\mbox{$\sigma_{\mathrm{Cs^{174}Yb}} = \left(5 \pm 2\right) \times 10^{-13} \,
\mathrm{cm^{2}}$} and $\dot{T}_{\mathrm{Cs,Heat}} = 4 \pm 1 \, \mathrm{\mu K /
s}$, describes the dynamics of the system well. The large fractional
uncertainty in the value of the elastic cross section is primarily due to the
large uncertainty in the spatial overlap. We have investigated the effect
of systematic errors in the measured parameters of our model and found that the
uncertainty in the trap frequency is dominant, and is larger than the
statistical error. Inclusion of the correction of (\ref{eq:correction}) is
important because the weaker confinement of Yb produces a vertical separation
between the two species, reducing the spatial overlap. Initially $F_{z}
\left(\Delta z\right)\sim0.75$. Over the timescale of the measurement the
spatial overlap reduces further due to the decreasing width of the Cs cloud as
it cools. The final value of $F_{z} \left(\Delta z\right)\sim0.6$.

Although the total heating rate is large, $\dot{T}_{\mathrm{Cs,Heat}} = 4 \pm 1
\, \mathrm{\mu K / s}$, it results from the sum of two heating mechanisms,
recombination heating and heating from the optical potential. The value for
recombination heating is reasonable because the Cs trap depth of \mbox{85
$\mu$K} is large enough to trap some of the products of the three-body
recombination event. For our scattering length, $a_{\mathrm{CsCs}} \approx 250
\, a_{0}$, $T_{\mathrm{Cs,H}}$ is still within the range from $2\epsilon/9$ to
$\epsilon/3$ proposed by the simple model in Ref.\ \cite{Weber2003}, where
$\epsilon = \hbar^{2}/m_{\mathrm{Cs}}\left(a_{\mathrm{CsCs}}-
\bar{a}\right)^{2}$ with $\bar{a} = 95.5 \, a_{0} $ for Cs. We also cannot rule
out any heating effects due to the broadband, multi-mode nature of the trapping
laser \cite{Sofikitis2011,Menegatti2013,Lauber2011,Yamashita2017} which may inflate the value of
$\dot{T}_{\mathrm{Cs,Heat}}$ above the simple estimate of $\mathrm{60 \, nK/s}$
based upon off-resonant scattering of photons. We find that varying the value
of the total trap heating rate $\dot{T}_{\mathrm{Cs,Heat}}$ over a large range
changes the extracted cross section by less than its error.

\begin{figure}
	\centering
		\includegraphics[width=1\linewidth]{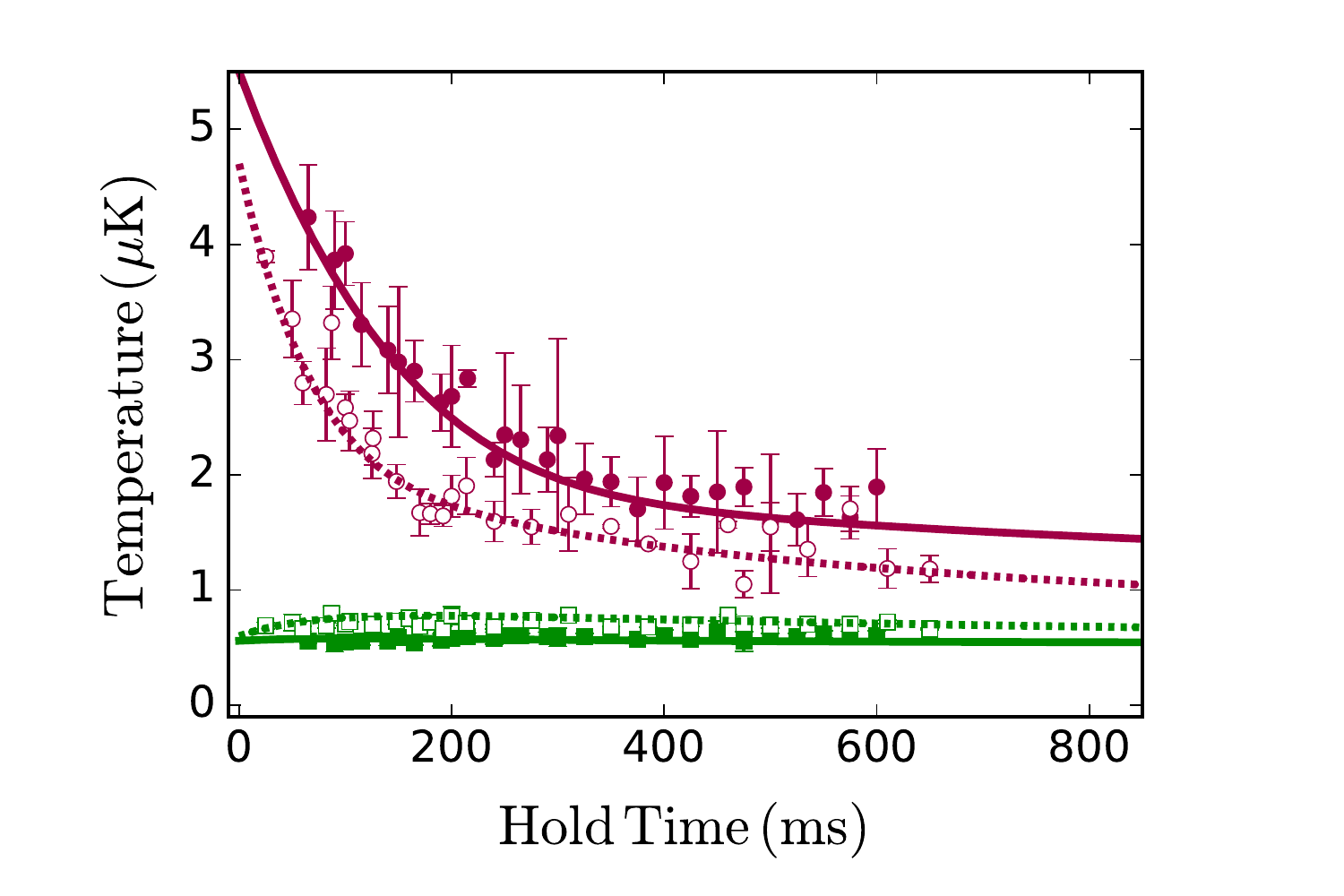}
\caption{Thermalization measurements of Cs (red circles) and Yb (green squares)
as a function of hold time in the ODT with the other species present. The
filled symbols are for $^{174}$Yb as coolant and open symbols are for
$^{170}$Yb as coolant. The solid (dotted) lines shows the best fit of our
model with $^{174}$Yb ($^{170}$Yb) as coolant.}
	\label{fig:TwoIsotopes}
\end{figure}

For the measurements presented in \mbox{Fig.\ \ref{fig:CsYb}}, we deliberately
use a low initial density of Cs atoms to avoid three-body recombination
collisions dominating the thermalization. This necessitates use of the weakest
possible trap and restricts the number of Cs atoms to $9 \times 10^{4}$.
However, due to the large ratio of polarizabilities between Cs and Yb (and the
effect of gravity), this results in a very shallow trap for Yb. Preparation of
Yb atoms in this shallow trap requires that the intraspecies scattering length
be favorable for evaporation, currently limiting the Yb isotopes we can study
to $^{170}$Yb and $^{174}$Yb. In \mbox{Fig.\ \ref{fig:TwoIsotopes}} we present
our thermalization measurements for $^{170}$Yb alongside those for $^{174}$Yb.
From the fit to the temperature profile we extract an effective cross section
\mbox{$\sigma_{\mathrm{Cs^{170}Yb}} = \left(18 \pm 8\right) \times 10^{-13} \,
\mathrm{cm^{2}}$} and $\dot{T}_{\mathrm{Cs,Heat}} = 5 \pm 2 \, \mathrm{\mu K /
s}$. The larger interspecies cross section allows Cs to be cooled to a lower
equilibrium temperature than with $^{174}$Yb. Due to the difference in the natural abundance (31.8\%
for $^{174}$Yb and 3.0\% for $^{170}$Yb \cite{Laeter2009}) and the intraspecies
scattering lengths ($a_{174} = 105 \, a_{0}$ and $a_{170} = 64 \, a_{0}$
\cite{Kitagawa2008}), we obtain a number of $^{170}$Yb atoms that is half that
of $^{174}$Yb, leading to a greater final temperature for $^{170}$Yb.

{\section{Calculated cross sections and extraction of scattering lengths}\label{sec:ScattTheory}

\begin{figure*}
	\centering
		\includegraphics[width=1\linewidth]{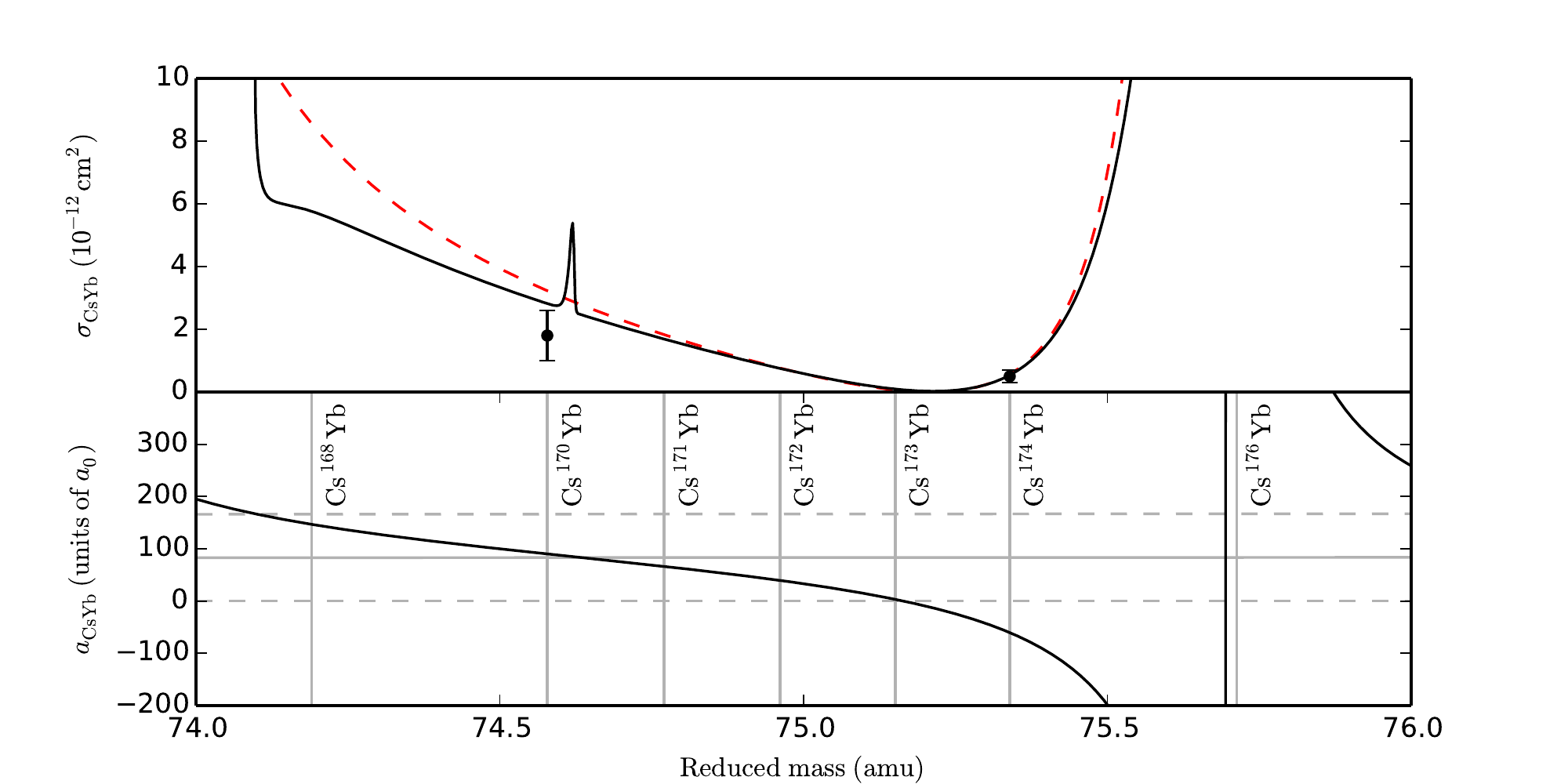} \caption{Top
panel: Thermalization cross sections as a function of reduced mass, calculated
on potentials optimized using the thermally averaged $\sigma_\eta^{(1)}$ given
by Eq. \ref{eq:sigma_eff} (black solid line) and the approximation $\sigma=4\pi
a^2$ (red dashed line). Points show experimentally measured cross sections and
error bars correspond to 1 standard deviation. Bottom panel: Calculated
scattering length as a function of reduced mass for the potential optimized
using $\sigma_\eta^{(1)}$. Vertical lines correspond to stable isotopes of Yb.
Horizontal lines correspond to 0, $\bar{a}$ and $2\bar{a}$. }
	\label{fig:mass_scaling}
\end{figure*}

Except near narrow Feshbach resonances, CsYb collisions can be treated as those
of two structureless particles with an interaction potential $V(R)$, which
behaves at long range as $-C_6 R^{-6}$. The scattering length for such a system
may be related to $v_{\rm D}$, the non-integer vibrational quantum number at
dissociation, by
\begin{equation}
a = \bar{a}\left[1-
\tan\left(v_{\rm D}+\textstyle\frac{1}{2}\right)\pi\right],
\label{eq:avD}
\end{equation}
where $\bar{a} = 0.477 988\ldots (2\mu C_6/\hbar^2)^{-1/4}$ is the mean
scattering length \cite{Gribakin1993} and $\mu$ is the reduced mass. For CsYb,
with $\sim 70$ bound states \cite{Brue2013}, changes in the Yb isotope alter
$v_{\rm D}$ by less than 1, so that the scattering lengths for all possible
isotopologs may be placed on a single curve. Determination of the scattering
length for one isotopolog allows predictions for all others.

The connection between scattering lengths and effective cross sections for
thermalization may be made at various different levels of sophistication. At
the lowest temperatures, thermalization is governed by the elastic cross
section $\sigma_{\rm el} = 4\pi a^2$. Various energy-dependent corrections to
$\sigma_{\rm el}$ may be included, from effective-range effects or higher
partial waves. However, when higher partial waves contribute to the scattering,
it is important to replace $\sigma_{\rm el}$ with the transport cross section
$\sigma_\eta^{(1)}$, which accounts for the anisotropy of the differential
cross section \cite{Anderlini2006, Frye2014}. p-wave scattering contributes to
$\sigma_\eta^{(1)}$ at considerably lower energy than to $\sigma_{\rm el}$,
because of the presence of interference terms between s~waves and p~waves.

In the present work we calculate $\sigma_\eta^{(1)}$ explicitly from scattering
calculations as described in Ref.\ \cite{Frye2014}, using the CsYb interaction
potential of Ref.\ \cite{Brue2013}. The resulting energy-dependent cross
sections are thermally averaged \cite{Anderlini2006},
\begin{equation}
\sigma_{\rm CsYb}(T) = \frac{1}{2} \int_0^\infty x^2 \sigma_\eta^{(1)}(x) e^{-x}\,dx,
\label{eq:sigma_eff}
\end{equation}
where $x=E/k_{\rm B}T$ is a reduced collision energy. The thermal average is
performed at the temperature $T=\mu (T_\mathrm{Cs}/m_\mathrm{Cs} +
T_\mathrm{Yb}/m_\mathrm{Yb})$ that characterizes the relative velocity. Note
that Eq.\ \ref{eq:sigma_eff} contains an extra factor of $x$ because
higher-energy collisions transfer more energy for the same deflection angle.

The scattering length of Eq.\ \ref{eq:avD} depends only on the fractional part
of $v_{\rm D}$. Small changes in the integer part of $v_{\rm D}$ have little
effect on the quality of fit. We therefore choose to fit the experimental cross
sections by varying the interaction potential of Ref.\ \cite{Brue2013} by the
minimum amount needed, retaining the number of bound states. We vary the
magnitude of its short-range part by a factor $\lambda$ to vary the scattering
length, while keeping the long-range part $-C_6 R^{-6}$ fixed, with $C_6$ taken
from Ref.\ \cite{Brue2013}. We perform scattering calculations using the
MOLSCAT package \cite{molscat2011}, with the SBE post-processor \cite{SBE} to
evaluate $\sigma_\eta^{(1)}$ from S-matrix elements.

We obtain optimal values of the potential scaling factor $\lambda$ by
least-squares fitting to the experimental cross sections. The fit using cross
sections from Eq.\ \ref{eq:sigma_eff} is shown by the solid line in the upper
panel of Fig.\ \ref{fig:mass_scaling}. Also shown (dashed line) is a fit using
the approximation $\sigma=4\pi a^2$, with $a$ obtained from scattering
calculations using MOLSCAT. The full treatment of Eq.\ (\ref{eq:sigma_eff})
gives a better fit than the approximation, and it may be seen that there are
much larger deviations between the two approaches for other reduced masses.
Even though the temperatures of both species are well below the p-wave barrier
height of $\sim 40$ $\mu$K, there is still considerable tunneling through the
barrier, which makes important contributions to the thermalization cross
sections $\sigma_\eta^{(1)}$ because of the interference between s-wave and
p-wave scattering. The deviations between the approaches are particularly large
where $a$ is close to $2\bar{a}$, producing a p-wave shape resonance
\cite{Gao2000}, and there is also a sharp feature where $a$ is close to
$\bar{a}$, producing a d-wave shape resonance.

The scattering lengths predicted using the best fitted interaction potential
are shown in the lower panel of Fig.\ \ref{fig:mass_scaling} for all
isotopologs of CsYb. The statistical uncertainties in the scattering lengths
are quite small, $a = 90 \pm 2$ $a_0$ for Cs$^{170}$Yb and $-60 \pm 9$ $a_0$
for Cs$^{174}$Yb. The fractional error is smaller for $^{170}$Yb than for
$^{174}$Yb, because Cs$^{170}$Yb is in a region of reduced mass $\mu$ where $a$
varies only slowly with $\mu$ and is mostly determined by $\bar{a}$, which is
accurately known. The systematic uncertainties arising from errors in the
number of bound states and the kinetic modeling are harder to quantify, but the
qualitative features should nevertheless be reliable. The calculated
interspecies scattering lengths are moderately positive for Yb isotopes from
168 to 172, close to zero for 173, and moderately negative for 174. The
scattering length for Cs$^{176}$Yb is predicted to be very large, which may
produce relatively broad Feshbach resonances \cite{Brue2013}.

\begin{figure}
	\centering
		\includegraphics[width=0.95\linewidth]{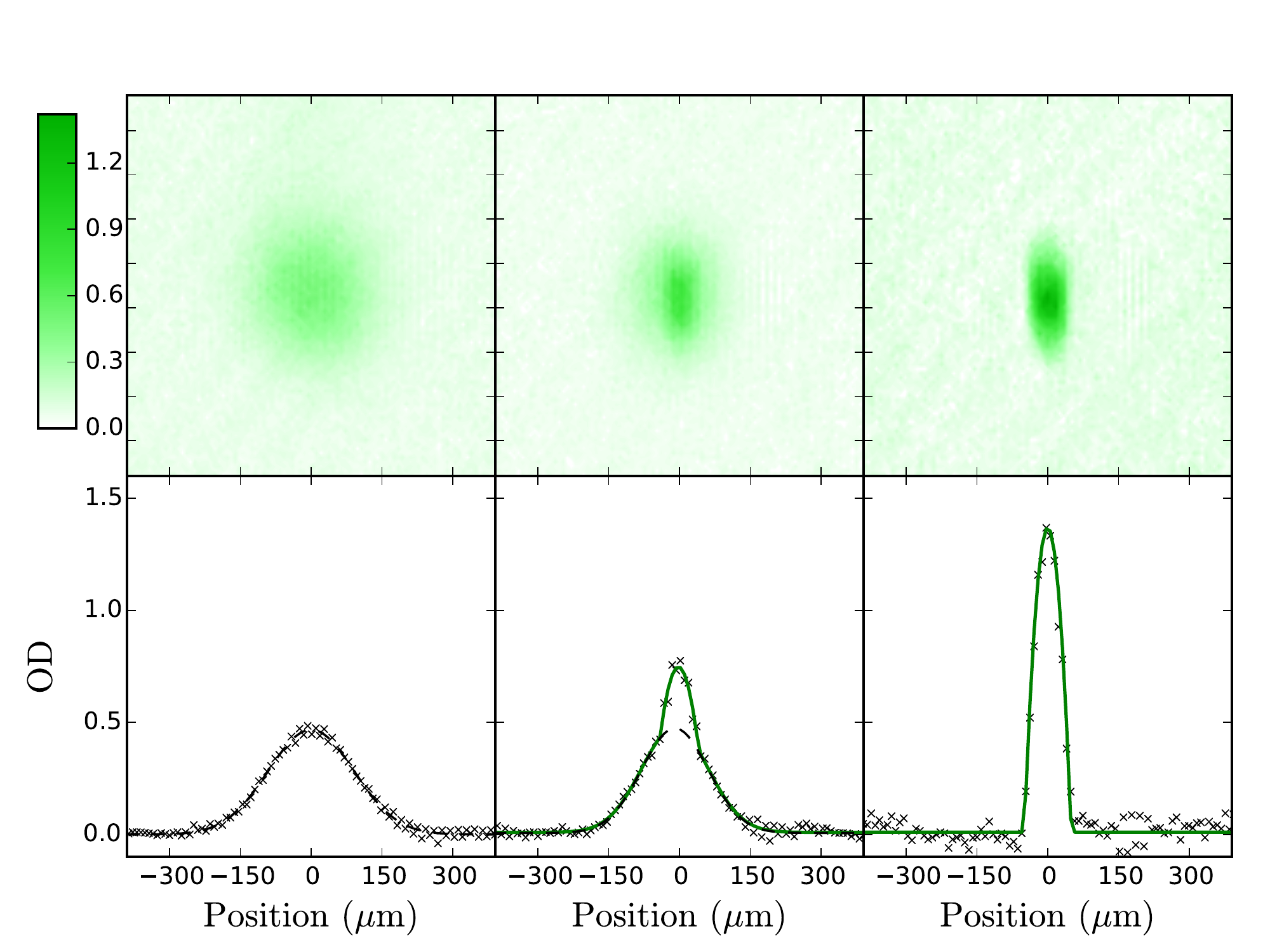}
\caption{Bose-Einstein condensation of $^{174}$Yb by evaporative cooling. The
top panels show absorption images taken after a \mbox{25 ms} time of flight.
The bottom panels show the corresponding horizontal cross-cuts through the
images. The laser power of the optical dipole trap is gradually reduced,
cooling the atoms from a thermal cloud at $T_{\mathrm{Yb}}=500\,$nK (left),
across the BEC transition to $T_{\mathrm{Yb}}=300\,$nK (middle). Finally, at
the end of evaporation a pure condensate (right) is produced containing $3
\times 10^{5}\,$atoms.}
	\label{fig:YbBEC}
\end{figure}

\section{Towards a doubly degenerate mixture}

In our current system, we can independently create Bose-Einstein condensates of
Cs and the two Yb isotopes with positive intraspecies scattering lengths and
workable abundance, $^{174}$Yb and $^{170}$Yb. The creation of a miscible
two-species condensate requires that \mbox{$g_{\mathrm{CsYb}}^{2} <
g_{\mathrm{YbYb}}\, g_{\mathrm{CsCs}}$}, where the interaction coupling
constants are \cite{Riboli2002}
\begin{equation}
g_{ij} = 2\pi\hbar^{2}a_{ij}\left(\frac{m_{i}+m_{j}}{m_{i}m_{j}}\right).
\end{equation}
The interspecies scattering lengths obtained above show that both Cs+$^{174}$Yb
and Cs+$^{170}$Yb BEC mixtures will be miscible at the magnetic field required
to minimize the Cs three-body loss rate. Moreover, Fig.~\ref{fig:mass_scaling}
shows that the interspecies scattering length is predicted to be of moderate
magnitude $(<200\,a_0)$ for all Yb isotopes except $^{176}$Yb. It should thus
also be possible to create stable, miscible quantum-degenerate Cs+Yb mixtures
for the less abundant $^{168}$Yb bosonic isotope and the two fermionic
isotopes, $^{171}$Yb and $^{173}$Yb. Note that the intraspecies scattering
length for $^{172}$Yb is large and negative \cite{Kitagawa2008}, precluding the
creation of a large condensate.

The realization of a doubly degenerate mixture will require a trapping
arrangement for both species that balances their individual requirements.
As a first step, we have created individual BECs of $^{174}$Yb
and Cs in the same apparatus. The differing requirements of the two species
are highlighted by the different routes we use to create the two
condensates. For $^{174}$Yb, the initial steps are the same as in preparing
the $^{174}$Yb gas for the thermalization measurements (see \mbox{Fig.\
\ref{fig:sequence}}), where the sample is loaded into the ODT at high power and
evaporatively cooled. However, for BEC the evaporation is continued until the
temperature is below the critical temperature, $T_{\mathrm{c,Yb}} \approx 350
\, \mathrm{nK}$, with the power in the ODT ramped down to around \mbox{400 mW}.
\mbox{Figure \ref{fig:YbBEC}} shows the transition to BEC for $^{174}$Yb. We
typically produce pure $^{174}$Yb BECs containing 3 to $4 \times 10^{5}$ atoms.

\begin{figure}
	\centering
		\includegraphics[width=0.95\linewidth]{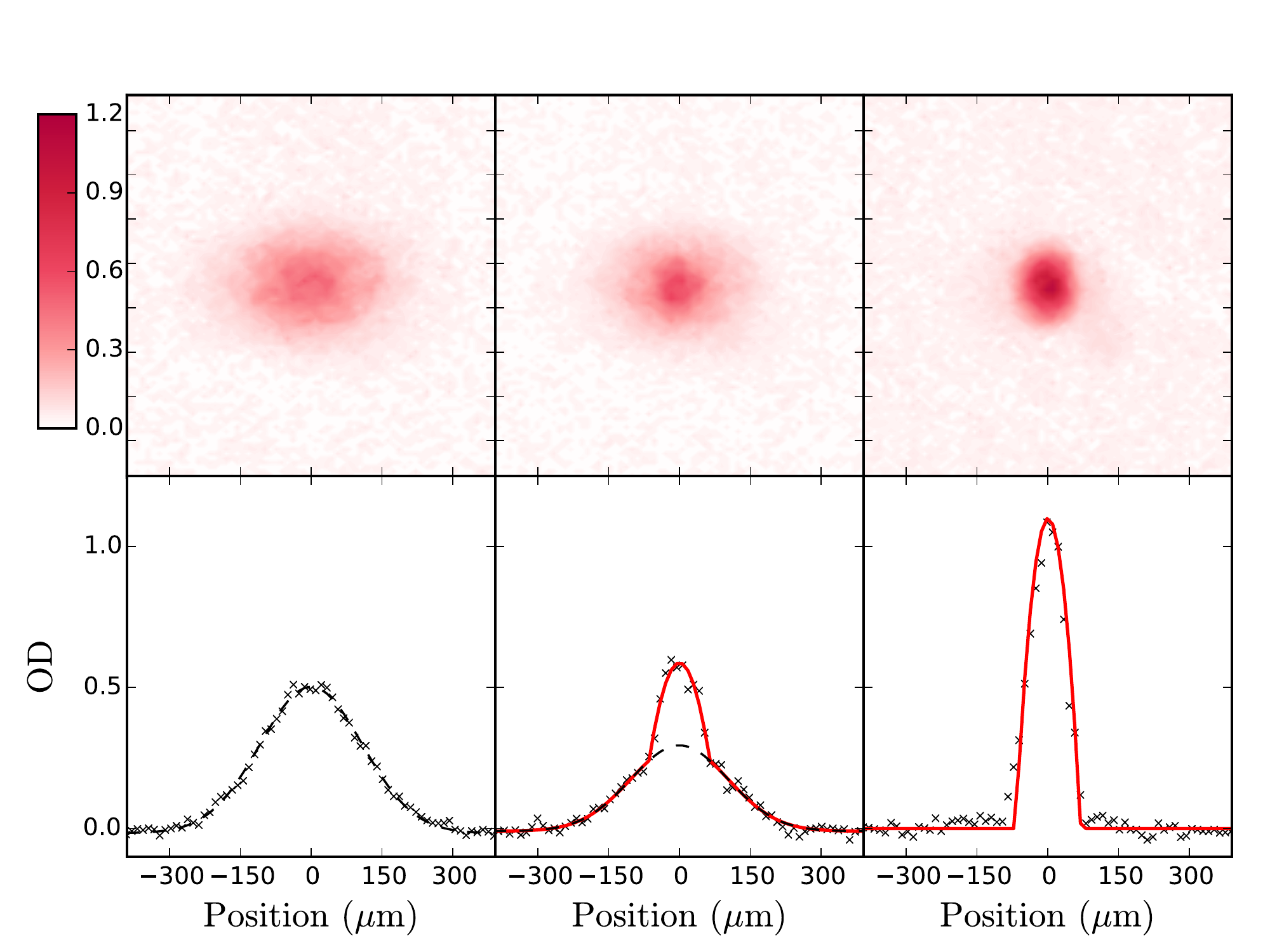}
\caption{Bose-Einstein condensation of Cs by evaporative cooling. The
top panels show absorption images taken after \mbox{50 ms} of levitated time of
flight. The bottom panels show the corresponding horizontal cross-cuts through
the images. The laser power of the optical dipole trap is gradually reduced,
cooling the atoms from a thermal cloud at $T_{\mathrm{Cs}}=70\,$nK (left),
across the BEC transition to $T_{\mathrm{Cs}}=50\,$nK (middle). Finally, at the
end of evaporation a pure condensate (right) is produced containing $5 \times
10^{4}\,$atoms.}
	\label{fig:CsBEC}
\end{figure}

Preparation of a Cs BEC is a contrasting case. The scattering properties of Cs
make the evaporative cooling approach used for $^{174}$Yb impractical. We
follow the approach of Ref.\ \cite{Kraemer2004} and use DRSC to pre-cool a
sample of $5 \times 10^{7}$ atoms to a temperature $T_{\mathrm{Cs}} = 2 \,
\mathrm{\mu K}$. These atoms are then transferred into a large-volume crossed
dipole trap which we call the reservoir trap (see Fig.\ \ref{fig:optics}). The
reservoir trap is created using a $50 \,$W fiber laser (IPG YLR-50-LP)
operating at $1070 \pm 3 \,$nm and is formed by two $20 \,$W beams crossing at
an angle of $25^{\circ} $, with waists $440 \pm 10 \, \mathrm{\mu m}$ and $640
\pm 20 \, \mathrm{\mu m}$. We initially load $1.5 \times 10^{7}$
atoms into the reservoir at $T_{\mathrm{Cs}}= 2.3 \, \mathrm{\mu K}$. We then
transfer $9 \times 10^{5}$ of these atoms into the ODT used previously for
thermalization measurements (and for $^{174}$Yb BEC); this is also known as the
dimple trap. The bias field is then reduced to $22.3 \,$G to optimize the ratio
of elastic to three-body recombination collisions \cite{Weber2003a} and the
sample is evaporatively cooled by reducing the dimple laser power over
$2.5\,$s. The onset of degeneracy occurs at $T_{\mathrm{c,Cs}} \approx 60 \,
\mathrm{nK}$, with the power in the dimple reduced to just $3 \,$mW. The BEC
transition for Cs is shown in \mbox{Fig.\ \ref{fig:CsBEC}}. We typically obtain
pure condensates of 4 to $5 \times 10^{4}$ atoms.

Simultaneously creating Cs and Yb BECs in the same optical dipole trap will be
very challenging. Table \ref{table} illustrates that the final trapping powers
needed for the independent creation of Cs and Yb BECs are very different. The
final trap for $^{174}$Yb requires $400 \,$mW of optical power at $1070\,$nm,
whereas that for Cs requires just $3 \,$mW, even though the trap depths differ
by only a factor of 5. The large difference in power stems from the combined
effects of the large ratio of the polarizabilities at $1070\,$nm
($\alpha_{\mathrm{Cs}}/\alpha_{\mathrm{Yb}}= 7.2$) and gravity reducing the
depth of the Yb potential. Additionally, the two species require very different
traps for efficient evaporative cooling. Cs demands careful management of the
three-body loss rate throughout evaporation, requiring more relaxed trapping
frequencies and necessitating the use of a reservoir and a dimple trap. By
contrast, $^{174}$Yb evaporates most efficiently in a tight trap with large
trapping frequencies, due to its much lower three-body loss coefficient
\cite{Takasu2003}. These differences suggest that double degeneracy is
unfeasible in the current trapping arrangement.

\begin{table}
	\begin{ruledtabular}
	\begin{tabular}{@{}l l l@{}}
	
	Parameter  & $^{174}$Yb & Cs \\
	\hline
	$U / k_{\mathrm{B}} \, \left(\mu \mathrm{K}\right)$ & 2.5 & 0.5 \\
	$\bar{\omega} / 2\pi$ (Hz)& 110 & 45 \\
	$\mathrm{P}_{\mathrm{Dimple}}$ (mW)& 400 & 3 \\
	$T_{\mathrm{c}} \, \left(\mathrm{nK}\right)$ & 350 & 60 \\
	$N \, \left(T_{\mathrm{c}}\right)$ & $9 \times 10^{5}$ & $8 \times 10^{4}$ \\
	\end{tabular}
	\end{ruledtabular}
	\caption{Parameters for Cs and $^{174}$Yb at the BEC
transition.}\label{table}
\end{table}

One solution is to create a tunable optical dipole trap for Cs using a trap
wavelength between the Cs $D_{1}$ and $D_{2}$ lines. In this region the Cs
polarizability varies from large positive to large negative values while the Yb
polarizability remains almost unchanged. This will allow the ratio of the trap
depths (and trap frequencies) to be tuned to a value more favorable for
simultaneous cooling of both species. However, due to the relatively low
detuning from atomic resonances, photon scattering may become a significant
issue for Cs during evaporation. Nevertheless, such a trap could be used to
combine two BECs after independent evaporation.

An alternative approach is the use of a bichromatic trap at $532 \,$nm and
$1064 \,$nm. The negative polarizability of Cs at $532 \,$nm
($\alpha_{\mathrm{Cs}} \left(532 \, \mathrm{nm}\right) = -210 \, a_{0}^{3}$)
balances the large positive polarizability at $1070 \,$nm
($\alpha_{\mathrm{Cs}}\left(1070 \, \mathrm{nm}\right) = 1140 \, a_{0}^{3}$).
Because the dominant transition of Yb is at $\lambda = 399 \,$nm, Yb is trapped
at both wavelengths ($\alpha_{\mathrm{Yb}}\left(532 \, \mathrm{nm}\right) = 240
\, a_{0}^{3}$, $\alpha_{\mathrm{Yb}}\left(1070 \, \mathrm{nm}\right) = 150 \,
a_{0}^{3}$). Tuning the power of the $532 \,$nm trap lasers thus allows the
ratio of trap depths for the two species to be tuned to a suitable value
\cite{Tassy2010,Vaidya2015}.

\section{Conclusions}

We have measured thermalization in an ultracold mixture of Cs and Yb. We have
used a kinetic model to determine the cross sections for interspecies
thermalization, taking account of additional heating effects that prevent
complete thermalization of the two species. We obtain values of
\mbox{$\sigma_{\mathrm{Cs^{174}Yb}} = \left(5 \pm 2\right) \times 10^{-13} \,
\mathrm{cm^{2}}$} and \mbox{$\sigma_{\mathrm{Cs^{170}Yb}} = \left(18 \pm
8\right) \times 10^{-13} \, \mathrm{cm^{2}}$}. We have carried out quantum
scattering calculations of the thermalization cross sections, taking account of
anisotropic scattering and thermal averaging, and fitted the short-range part
of the CsYb interaction potential to reproduce the experimental results. We
have used the resulting interaction potential to calculate scattering lengths
for all isotopologs of CsYb.

The interspecies Cs+Yb scattering lengths are predicted to have moderate
magnitudes $(<200\,a_0)$ for all Yb isotopes except $^{176}$Yb, with good
prospects of creating doubly degenerate mixtures. We have cooled
both $^{174}$Yb and Cs to degeneracy in the same apparatus, but cooling both
species to degeneracy in the same optical trap will be challenging, as
illustrated by the contrasting routines we use to produce independent BECs of
the two species. We have discussed the use of a tunable-wavelength or
bichromatic optical trap that should allow co-trapping of quantum-degenerate
Cs+Yb mixtures.

The optimized CsYb potential will assist direct measurements of the CsYb
binding energies using two-photon photoassociation spectroscopy
\cite{Kitagawa2008,Munchow2011}. Precise determination of the near-threshold
bound states of the molecular potential will allow the accurate prediction and
subsequent experimental search for Feshbach resonances that can be used in
magnetoassociation to form ultracold $^{2}\Sigma$ CsYb molecules.

The data presented in this paper are available online \footnote{Durham University Collections, \url{http://dx.doi.org/10.15128/r2rv042t06m}}.

\begin{acknowledgments}
We thank I. G. Hughes, M. R. Tarbutt and E. A. Hinds for many valuable
discussions. We acknowledge support from the UK Engineering and Physical
Sciences Research Council (Grant Nos. EP/I012044/1 and EP/P01058X/1).
\end{acknowledgments}

\appendix*
\section{Modeling Thermalization}

In order to simulate sympathetic cooling of two distinct atomic species, a
simple kinetic model can be used. Here we will derive a system of 4 coupled
equations which describe the evolution of the number ($N_{\mathrm{Cs}},
N_{\mathrm{Yb}}$) and temperature ($T_{\mathrm{Cs}}, T_{\mathrm{Yb}}$) of the
two species. These can then be solved numerically for given initial conditions
and be compared with experimental results (see, for example, Fig.\
\ref{fig:CsYb}). Differentials with respect to time are denoted with a dot
above the symbol, for example $\dot{N}$. Here we draw together different
elements from several similar models
\cite{Luiten1996,Ketterle1996,Mosk2001,OHara2001,Weber2003,Anderlini2005,Tassy2010,Ivanov2011}.

We first consider single-species effects, starting with evaporative cooling. If
the temperature of the species is not far enough below the trap depth $U_{i}$
then atoms with sufficient energy may evaporate from the trap. The
dimensionless parameter $\eta_i=U_{i}/k_{\mathrm{B}}T_{i}$ characterizes the
trap depth relative to the temperature. Assuming that $k_{\mathrm{B}}T_{i}$ is
small compared to $U_{i}$, atoms with energy greater than the trap depth are
produced at a rate of $\Gamma_{ii}\eta_i \exp(-\eta_i)$
\cite{Luiten1996,Ketterle1996}. The total effective hard-sphere elastic
collision rate $\Gamma_{ii}$ is conveniently written as $N_i\gamma_{ii}$ where
the effective mean collision rate per atom is
\begin{equation}
\gamma_{ii}=\left<n_{i}\right>_{\rm sp}\sigma_{ii}\bar{v}_{ii},
\end{equation}
where $\sigma_{ii}$ is the elastic scattering cross-section, $\bar{v}_{ii}=\sqrt{16
k_{\mathrm{B}} T_i/ \pi m_i}$ is the mean velocity, $m_i$ is the mass of species $i$, and the mean density is given by
\begin{equation}
\left<n_{i}\right>_{\rm sp}=\frac{N_i}{8}\left(\frac{m_i \overline{\omega}_i^{2}}{\pi k_{\mathrm{B}}T_i}\right)^{3/2},
\end{equation}
where $\overline{\omega}_i=\sqrt[3]{\omega_{x}\omega_{y}\omega_{z}}$ is the
mean trap frequency. When an atom evaporates from the trap, it carries away an
average energy $\epsilon_{\mathrm{evap}}=(\eta_i + \kappa_i)k_{\mathrm{B}}T_{i}$,
where $ \kappa_i = (\eta_i-5)/(\eta_i-4)$ \cite{OHara2001}; this expression for
$\kappa_i$ is appropriate if $k_{\mathrm{B}}T_{i}\ll U_{i}$ and $U_{i}$ is
harmonic near the minimum, as is the case in our experiment. The evolution of
number and total energy of the ensemble due to evaporation is therefore
\begin{align}
\dot{N}_{i,\mathrm{evap}}&=-N_i\gamma_{ii}\eta_i\exp(-\eta_i), \\
\dot{E}_{i,\mathrm{evap}}&=(\eta_i+\kappa_i)k_{\mathrm{B}}T_i\dot{N}_{i,\mathrm{evap}}.
\end{align}
The evolution of the temperature can be derived from the relation
$3 k_{\mathrm{B}}\left(T_{i}\dot{N_{i}}+\dot{T}_{i}N_{i}\right)=\dot{E}_{i}$, to give
\begin{equation}
\dot{T}_{i,\mathrm{evap}}=\eta_i \exp\left(-\eta_i \right)\gamma_{ii}
\left(1-\frac{\eta_i + \kappa_i}{3}\right)T_{i}.
\end{equation}

If present, inelastic or reactive two-body collisions could also be included in
this model. However, since both Cs and Yb are in their absolute ground state,
two-body collisional losses are fully suppressed for our case.

The effect of collisions with background gas is included through the terms
\begin{align}
\dot{N}_{i,\mathrm{bg}}&=-K_{\mathrm{bg}}N_i, \\
\dot{E}_{i,\mathrm{bg}}&=3k_{\mathrm{B}}T_i\dot{N}_{i,\mathrm{bg}},
\end{align}
where $K_{\mathrm{bg}}$ is the background loss rate, which is taken to be the
same for both species. There is no corresponding change in temperature as the
loss does not preferentially affect either warmer or cooler atoms.

The inclusion of three-body collisions is essential for this system, due to the
large three-body loss coefficient in Cs. The loss rate is given by
\begin{equation}
\dot{N}_{i,\mathrm{3}}=-K_{i,\mathrm{3}} \left< n_i^{2}\right>_{\rm sp} N_i,
\end{equation}
where $\left<n_i^2 \right>_{\rm sp}=\sqrt{64/27}\left<n_i \right>_{\rm sp}^2$
and $K_{i,3}$ is the three-body loss coefficient. Because of the density
dependence of three-body collisions, atoms are preferentially lost from the
high-density region near the center of the trap. The potential energy in this
region is lower than the ensemble average, meaning an average excess energy of
$1 \,k_{\mathrm{B}}T$ remains in the trap for each atom that is lost
\cite{Weber2003}. In addition to this effect, when three-body recombination
produces an atom and a diatom in a state very near threshold, the energy
released may be small enough that the atom is not lost, but remains trapped
along with a fraction of the energy released \cite{Weber2003}. This heating
contributes $k_{\mathrm{B}}T_{i,\mathrm{H}}$ per lost atom. The combination of
these two effects gives
\begin{align}
\dot{E}_{i,\mathrm{3}}&=\left(2 T_i- T_{i,\mathrm{H}}\right) k_{\mathrm{B}}\dot{N}_{i,\mathrm{3}}, \\
\dot{T}_{i,\mathrm{3}}&=K_{i,\mathrm{3}} \left< n_i^{2}\right>_{\rm sp} \frac{T_{i}+T_{i,\mathrm{H}}}{3}.
\end{align}

We also introduce an additional term $\dot{T}_{i,\mathrm{ODT}}$ to
account for extra heating from the trapping potential.

We now consider interspecies collisions and the thermalization they cause, as
modeling these is our primary purpose. The average energy transfer in a
hard-sphere collision is \cite{Mosk2001}
\begin{equation}
\Delta E_{\mathrm{Cs} \rightarrow \mathrm{Yb}} = \xi k_{\mathrm{B}}\Delta T,
\label{App_Eq:Delta_E}
\end{equation}
where $\Delta T=T_{\mathrm{Cs}}-T_{\mathrm{Yb}}$ and
\begin{equation}
\xi = \frac{4m_{\mathrm{Cs}}m_{\mathrm{Yb}}}{\left(m_{\mathrm{Cs}}+m_{\mathrm{Yb}}\right)^{2}}
\end{equation}
reduces the energy transfer for collisions between atoms of different masses.
If the collisions are not classical hard-sphere (or purely s-wave) in nature,
then different deflection angles $\Theta$ should be weighted by a factor of
$1-\cos\Theta$ \cite{Anderlini2006, Frye2014} and the average energy
transferred per collision varies from Eq.\ \ref{App_Eq:Delta_E}. Such effects
are not included explicitly in this simple kinetic treatment, so the resulting
cross sections and collision rates should be interpreted as \emph{effective
hard-sphere} quantities. We include the effects of deflection angles when we
calculate thermalization cross sections from scattering theory in Sec.\
\ref{sec:ScattTheory}.

In the hard-sphere model, the total energy transferred is just the average
energy transferred in a hard-sphere collision multiplied by an effective
hard-sphere collision rate $\Gamma_{\mathrm{CsYb}}$, giving
\begin{align}
\dot{E}_{\mathrm{Cs, therm}}&=-\xi k_{\mathrm{B}}\Gamma_{\mathrm{CsYb}} \Delta T, \\
\dot{E}_{\mathrm{Yb,therm}}&=+\xi k_{\mathrm{B}}\Gamma_{\mathrm{CsYb}} \Delta T,
\end{align}
and
\begin{align}
\dot{T}_{\mathrm{Cs,therm}}&=-\frac{\xi \Gamma_{\mathrm{CsYb}} \Delta T}{3N_\mathrm{Cs}}, \\
\dot{T}_{\mathrm{Yb,therm}}&=+\frac{\xi \Gamma_{\mathrm{CsYb}} \Delta T}{3N_\mathrm{Yb}}.
\end{align}
Since thermalization collisions do not produce loss,
$\dot{N}_{i,\mathrm{therm}}=0$. We can relate the effective rate to an
effective cross section $\sigma_{\mathrm{CsYb}}$ through the relation
$\Gamma_{\mathrm{CsYb}}=\bar{n}_{\mathrm{CsYb}}\sigma_{\mathrm{CsYb}}\bar{v}_{\mathrm{CsYb}}$.
Here,
\begin{equation}
\bar{v}_{\mathrm{CsYb}}=\sqrt{\frac{8 k_{\mathrm{B}}}{\pi}\left(\frac{T_{\mathrm{Yb}}}{m_{\mathrm{Yb}}}+\frac{T_{\mathrm{Cs}}}{m_{\mathrm{Cs}}}\right)}
\end{equation}
is the mean collision velocity.
The spatial overlap $\bar{n}_{\mathrm{CsYb}}$ is found by integrating the density distributions
of the two species
\begin{widetext}
\begin{align}
\bar{n}_{\mathrm{CsYb}}&=\int \left[n_{\mathrm{Cs, single}}\left(\bf{r}\right)+n_{\mathrm{Cs,cross}}\left(\bf{r}\right)\right]n_{\mathrm{Yb}}\left(\bf{r}\right) d^{3}r \nonumber \\
&=N_{\mathrm{Yb}}\frac{m_{\mathrm{Yb}}^{3/2} \overline{\omega}_{\mathrm{Yb}}^{3}}{2 \pi k_{\mathrm{B}}}\left[\frac{N_{\mathrm{Cs,single}}}{\left(T_{\mathrm{Yb}}+\beta_{\mathrm{single}}^{-2} T_{\mathrm{Cs}}\right)^{3/2}}+\frac{N_{\mathrm{Cs,cross}}}{\left(T_{\mathrm{Yb}}+\beta_{\mathrm{cross}}^{-2} T_{\mathrm{Cs}}\right)^{3/2}}\right], \label{App_Eq:interspecies_density}
\end{align}
\end{widetext}
where $\beta_{j}^2
=m_{\mathrm{Cs}}\overline{\omega}_{\mathrm{Cs},j}^{2}/m_{\mathrm{Yb}}\overline{\omega}_{\mathrm{Yb}}^{2}$,
where $j=\left\{\mathrm{single,cross}\right\}$ denotes the different cases for
Cs atoms trapped in the crossed- and single-beam regions. Eq.\
\ref{App_Eq:interspecies_density} holds true for two clouds centered at the same
position, but if the positions of the two clouds are offset in the $z$
direction by $\Delta z$
 then the spatial overlap must be reduced by a factor
\begin{equation}
F_{z} \left(\Delta z\right) = \exp\left(-\frac{m_{\mathrm{Yb}}\omega_{\mathrm{Yb},z}^{2}\Delta z^{2}}{2 k_{\mathrm{B}} \left(T_{\mathrm{Yb}}+\beta^{-2}_{\rm cross} T_{\mathrm{Cs}}\right)}\right) .
\end{equation}
In our case, the displacement of the clouds is due to gravitational sag and
\begin{equation}
\Delta z = g\left(\frac{1}{\omega^{2}_{\mathrm{Cs,z}}}-\frac{1}{\omega^{2}_{\mathrm{Yb,z}}}\right),\label{App_Eq:sag}
\end{equation}
where $g$ is the acceleration due to gravity.

Combining all these contributions gives Eqs.\ \ref{Eq:Number} and \ref{Eq:Temp}.

\bibliography{ThermalisationPaperReferences}

\end{document}